\documentclass[pra,twocolumn]{revtex4}
\usepackage{graphicx}

\newcommand{\gate}[1]{\textsf{#1}}

\newlength{\placeimageheight}

\newcommand{\placeimage}[2]{
\setlength{\placeimageheight}{#1}
\raisebox{0.5ex}{\raisebox{-0.5\placeimageheight}{
	\includegraphics[height=\placeimageheight]{#2}}}
}

\begin{document}

\title{Natural two-qubit gate for quantum computation using 
       the $XY$ interaction}
\author{Norbert Schuch}
\email{norbert.schuch@mpq.mpg.de}
\author{Jens Siewert}
\email{jens.siewert@physik.uni-regensburg.de}
\affiliation{Institut f\"ur Theoretische Physik, 93040 Regensburg, Germany}

\begin{abstract}
The two-qubit interaction Hamiltonian of a given physical implementation
determines whether or not a two-qubit gate such as the \gate{CNOT} gate can 
be realized easily. It can be shown that, e.g., with the $XY$
interaction more than one two-qubit operation is required in order
to realize \gate{CNOT}. Here we propose a two-qubit gate for the $XY$
interaction which combines \gate{CNOT} with the \gate{SWAP} operation. 
By using this gate quantum circuits can be implemented efficiently,
even if only nearest-neighbor coupling between the qubits is
available.
\end{abstract}


\date{\today}

\maketitle

\section{\label{sec:introduction}Introduction}

It is well known that there are universal sets of quantum gates
which are sufficient to perform any unitary operation on an arbitrary
number of qubits.  For example, it has been shown that arbitrary 
one-bit operations together with a non-trivial two-qubit gate provide such a
set~\cite{brylinski:universal_gates,bremner:all_are_universal}. 
Clearly, in order to perform an arbitrary
computation with three or more qubits the computer needs to be capable of
performing two-qubit operations between arbitrary pairs of qubits.

One possible universal set of gates is the set of arbitrary 
one-bit operations (i.e.\ all unitary transformations of a single qubit) 
together with the two-bit controlled NOT gate
(\gate{CNOT})~\cite{barenco:univ_gates_f_qc}.
The \gate{CNOT} operation is especially interesting since, to a
certain extent, it can be treated as a ``classical'' gate originating
from classical reversible computation. Gates of this type are
important since there are schemes which can be transferred directly 
from classical reversible computation to quantum circuits.
Thus the \gate{CNOT} gate has become an ubiquitous
reference in the design of quantum circuits. 
Correspondingly, the universality 
of a certain hardware setup is often demonstrated by providing ways
how to obtain the \gate{CNOT} gate (as well as arbitrary
one-bit gates) by controlled manipulations of the system parameters.

Typically the formal solution for a computational task (e.g., a complex
operation on several qubits, an error correction protocol etc.) is
given in terms of a sequence of one-bit gates and \gate{CNOT} operations.
For practical reasons it is desirable to optimize these sequences
according to certain criteria such as, e.g., the number of operations.
On the one hand  this helps to do as many computational steps as possible 
within a finite decoherence time, and on the other hand it renders the 
computation more stable with respect to computational errors.

The existence of a formal solution, however, does not mean that it can
be implemented directly in a physical system. 
There are certain---rather general---hardware-related issues which 
have to be considered before.
In quantum circuits for complex computational tasks it is assumed that
two-qubit operations can be performed on \emph{any} pair of qubits, 
i.e., that each qubit can be coupled with any other qubit.
While for many practical implementations it is possible to 
couple each qubit to a few others 
it often appears 
difficult to realize a uniform (and tunable)
coupling for \emph{any} pair of qubits. 

From a formal point of view, this is not a problem. Even if there
is only nearest-neighbor coupling, two arbitrary
qubits can be brought to interaction by consecutively swapping one
of them
with its nearest neighbor until the two considered qubits are next
to each other. Then the two-bit operation is carried out and the
swapping is reversed.  
However, from a practical point of view
this workaround is rather unsatisfactory since it increases the number of
operations and hence the length of the sequence considerably.

Apart from the ``range of the interaction'' there is another 
important issue related to the coupling. 
It depends on the microscopic coupling Hamiltonian whether or
not it is easy to implement the \gate{CNOT} gate (in particular,
whether more than one ``elementary'' two-qubit gate is required
to realize \gate{CNOT}, see the discussion
below). This seems to favor certain kinds of couplings with respect
to others right from the start.

The aim of this work is to show that such hardware-related difficulties
in practice may be overcome by adapting the design of the quantum
circuits under consideration. 
While hardware properties seem to render the implementation more 
difficult we demonstrate that by combining just these properties 
an efficient implementation can be achieved.
The paper is organized as follows. After introducing notations
(Section~\ref{sec:definitions}),
we discuss possible interaction Hamiltonians for two qubits 
and the corresponding ``elementary gates'' (Section~\ref{sec:int_hamilts}).
In Section~\ref{sec:cns_gate} we consider the properties of the so-called 
\gate{iSWAP} gate 
in more detail. This gate appears to be a natural choice for 
the elementary two-qubit gate if the coupling is given by
the $XY$ interaction. Finally we present two examples how 
complex networks can be implemented efficiently for the case
that only nearest-neighbor $XY$ coupling is available
(Section~\ref{sec:examples}).

\section{\label{sec:definitions}Definitions and general assumptions}

To fix ideas, let us first introduce 
the notation which will be used throughout the paper.

Each line in a quantum circuit denotes a single qubit, the operations act
on these lines from the left to the  right. 
One-qubit gates are usually denoted by a single box 
with the name of the operation (i.e.\ the matrix $M$):
$$
\mbox{\placeimage{1.5em}{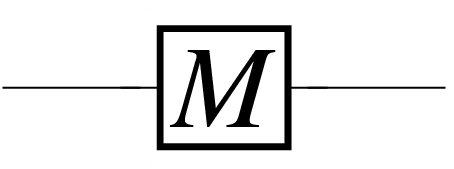}}\quad.
$$
Special one-bit gates are the rotations about the $x$ axis 
$$
R_x(\phi)=e^{-i\sigma_x\phi/2}=
    \left(\begin{array}{cc}\cos(\phi/2)&-i\sin(\phi/2)\\
	    -i\sin(\phi/2) & \cos(\phi/2)  \end{array}\right)
$$
and the $z$ axis
$$
R_z(\phi)=e^{-i\sigma_z\phi/2}=
\left(\begin{array}{cc}e^{-i\phi/2}&0\\0&e^{i\phi/2}\end{array}\right)\quad.
$$
These operations can be used to generate arbitrary one-bit operations; 
we denote them by \hspace{-1em}
\placeimage{1.5em}{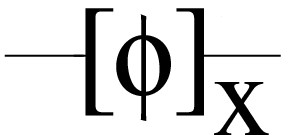} and \hspace{-1em}
\mbox{\placeimage{1.5em}{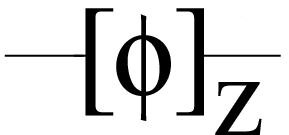}.}
The Hadamard
transformation $H$ can be built by using $x$ and $z$ rotations:
$$
\frac1{\sqrt{2}}\left(\begin{array}{cc}1&1\\1&-1\end{array}\right)
=
\!\!\mbox{\placeimage{1.5em}{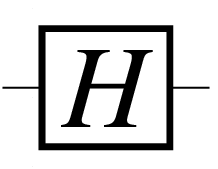}}=
\!\!\mbox{\placeimage{1.5em}{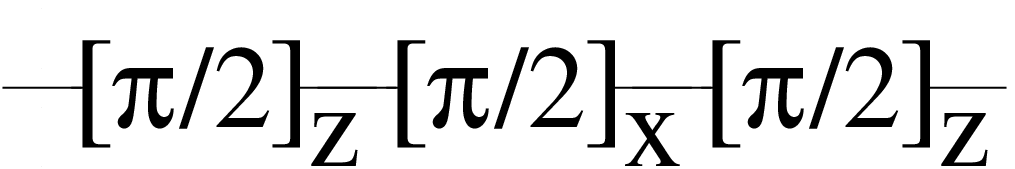}}\quad.
$$

The two-bit \gate{CNOT} operation is denoted by the 
following symbol
$$
\mbox{\placeimage{4em}{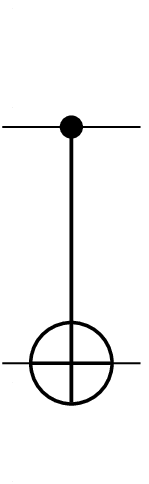}}=
\left(\begin{array}{cccc}
1\\&1\\&&0&1\\&&1&0
\end{array}\right)\quad.
$$
Here we use the standard (computational) basis for qubits, where the basis states
are ordered lexicographically, i.e.,
$\{|00\rangle,|01\rangle,|10\rangle,|11\rangle\}$.

Finally, the \gate{SWAP} gate exchanges the states of two qubits.
It can be obtained by using the \gate{CNOT} gate:
\begin{equation}
\label{eq:swap_cnot}
\mbox{\placeimage{4em}{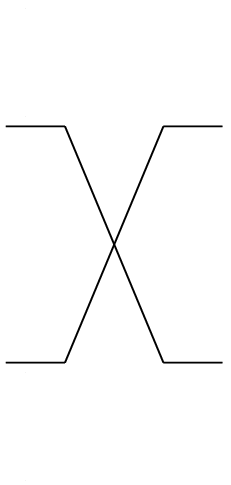}}=
\mbox{\placeimage{4em}{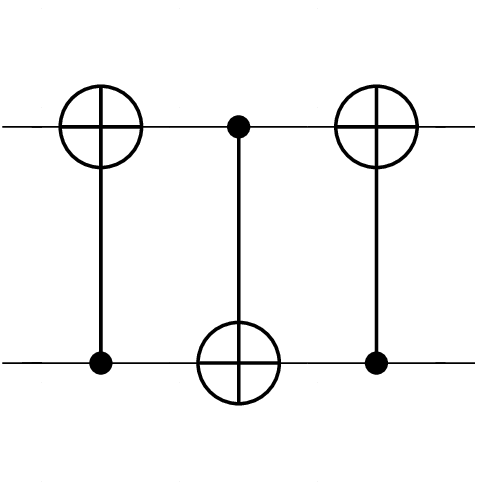}} \ \ .
\end{equation}

The Hamiltonian of a quantum computing device can be written as
$$
\mathcal H = \sum_{i=1}^N\mathcal H_i + \sum_{(i,j)} \mathcal H_{i,j}
$$
where $N$ is the number of qubits, the first term describes the single-qubit
parts $\mathcal H_i$ and the Hamiltonians $\mathcal H_{i,j}$ denote the interactions
between the qubits. The sum is taken over all pairs $(i,j)$ of qubits
which can be coupled at the hardware level. As we have mentioned before,
one would like to have interaction terms for arbitrary pairs of qubits. 
However, often this will be difficult to realize. 
For example, in many of the proposed solid-state implementations 
nearest-neighbor coupling appears to be the standard choice
(consider, e.g., electron spins in quantum dots coupled by tunnel 
junctions~\cite{loss:qc_w_qdots},  
excitonic qubits in quantum dots~\cite{rossi:excitons},
inductively coupled Josephson flux qubits~\cite{orlando:flux_qubits},
or Josephson charge qubits coupled by Josephson junctions~\cite{jens:JLTP}).

In the following we will assume that the $N$ qubits (on which 
arbitrary one-bit operations can be performed) are arranged in
a chain or in a ring (i.e., a chain with periodic boundary conditions)
with independently tunable nearest-neighbor interactions.
The premise of only nearest-neighbor coupling 
appears to be weak enough to render the discussion
sufficiently general.

\section{\label{sec:int_hamilts}Interaction Hamiltonians for qubits}

In this section we will illustrate that, depending
on the available coupling term $\mathcal H_{i,j}$, it is more
or less difficult to generate the \gate{CNOT} operation.
On the other hand, for each type of coupling there are two-qubit
gates which can be implemented in a  straightforward manner \emph{and}
which can be viewed as \emph{classical} gates (up to one-bit operations).

Although other choices for the interaction part of the Hamiltonian
are possible we will focus on three different types of couplings:
Firstly, the $ZZ$ interaction
$$\mathcal H^{ZZ}_{i,j}(E_{i,j}^{ZZ}) = 
		-\frac{E_{i,j}^{ZZ}}4\sigma^{(i)}_z\sigma^{(j)}_z\ ,
$$
which, e.g., can be realized 
for Josephson flux 
qubits~\cite{orlando:flux_qubits} or
for Josephson charge qubits coupled 
inductively~\cite{makhlin:jqubits_nature}. 
Secondly, the $JJ$ or Heisenberg interaction
$$
\mathcal H^{JJ}_{i,j}(E_{i,j}^{JJ}) 
    =-\frac{E_{i,j}^{JJ}}4\left[\sigma^{(i)}_x\sigma^{(j)}_x+ 
  \sigma^{(i)}_y\sigma^{(j)}_y+\sigma^{(i)}_z\sigma^{(j)}_z\right]\ ,
$$ 
which basically appears in systems where spins are coupled by
the exchange interaction, for example 
spins in quantum dots interacting via a tunnel junction~\cite{loss:qc_w_qdots}, nuclear spins in phosphorus-doped
silicon devices
\cite{kane:silic_nuclspin_qc}, or spin-resonance 
transistors~\cite{vrijen:spin_res_trans};
and finally the $XY$ interaction
$$
\mathcal H^{XY}_{i,j}(E_{i,j}^{XY}) 
	    =-\frac{E_{i,j}^{XY}}4\left[\sigma^{(i)}_x\sigma^{(j)}_x + 
	    \sigma^{(i)}_y\sigma^{(j)}_y\right] \ .
$$
This type of coupling has been proposed 
for quantum dot spins coupled by a cavity~\cite{imamoglu:pap_w_sqrtiswap},
for Josephson charge qubits coupled by Josephson junctions~\cite{jens:JLTP},
and for nuclear spins interacting via a two-dimensional electron 
gas~\cite{mozyrsky:nspin_2deg}.

For the $ZZ$-interaction, the \gate{CNOT} 
operation is indeed the natural two-bit
operation since 
$$
\exp\left[-i \mathcal H^{ZZ}_{i,j}(E_{i,j}^{ZZ}) 
\frac{\pi}{E_{i,j}^{ZZ}}\right]=
e^{i\pi/4}
\left(\begin{array}{cccc}
1\\&-i\\&&-i\\&&&1
\end{array}\right)
$$
is equvialent
to \gate{CNOT} 
up to one-bit operations~\cite{makhlin:2bit_equiv}. 
As opposed to this, the Heisenberg interaction does not yield the \gate{CNOT}
operation directly, while
$$
\exp\left[-i \mathcal H^{JJ}_{i,j}(E_{i,j}^{JJ})
\frac{\pi}{E_{i,j}^{JJ}}\right]=
e^{i\pi/4}
\left(\begin{array}{cccc}
1\\&0&1\\&1&0\\&&&1
\end{array}\right)
$$ 
corresponds to the \gate{SWAP} operation.
Since \gate{SWAP} cannot entangle two qubits one has to use alternative
ways to produce \gate{CNOT}, e.g., via the square root of \gate{SWAP}
(briefly denoted by $\sqrt{\mbox{\gate{SWAP}}}$)
which can be obtained by applying the Heisenberg Hamiltonian only for a
time $0.5\,\pi\hbar/E_{i,j}^{JJ}$. Then \gate{CNOT} can be 
generated~\cite{loss:qc_w_qdots} via 
$$
\mbox{\placeimage{3.8em}{cnot.eps}}=
$$\vspace{-1em}$$
\!\!\!\!\!\!\mbox{\placeimage{3.8em}{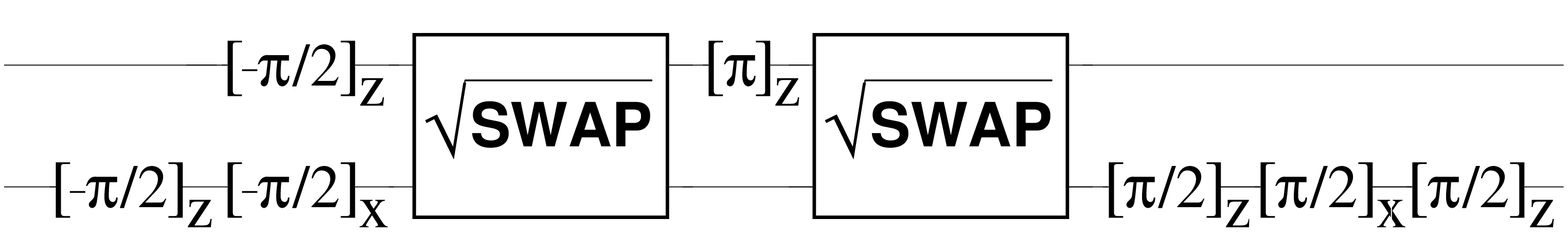}}
$$
However, here
$\sqrt{\mbox{\gate{SWAP}}}$ has to be applied twice: 
the \gate{CNOT} gate \emph{cannot} be
constructed by applying  a $\mathcal H^{JJ}_{i,j}$-based 
gate only once~\cite{makhlin:2bit_equiv}.
The converse is true for the construction of the \gate{SWAP}
operation using the $ZZ$ interaction: while \gate{CNOT} can be obtained in
one step, \gate{SWAP} requires two two-bit operations.

In Ref.~\cite{makhlin:2bit_equiv} it was also shown
that the $XY$ interaction Hamiltonian $\mathcal H^{XY}_{i,j}$ can \emph{neither}
generate the \gate{CNOT} operation \emph{nor} the \gate{SWAP} operation by
applying an $XY$-based gate only once.
Nevertheless it is sufficient to build
a \gate{CNOT} gate.
An appropriate ``elementary'' two-bit gate is the 
\gate{iSWAP} operation which is obtained by applying
$\mathcal H^{XY}_{i,j}(E_{i,j}^{XY})$ for a
time $t=\pi\hbar/E_{i,j}^{XY}$:  
$$
\!\!\!\!\!\mbox{\placeimage{4em}{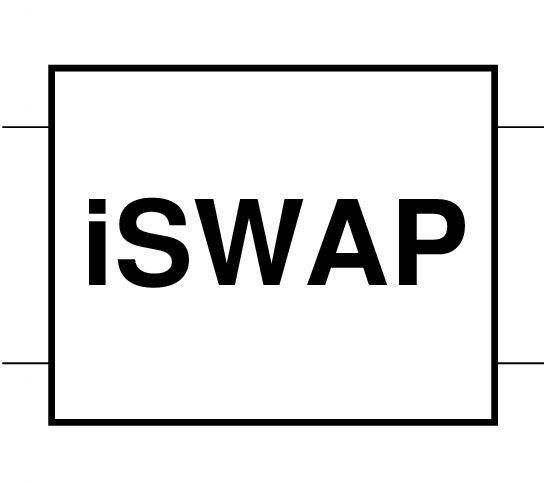}}\!\!\!:=\!\!
\left(\begin{array}{cccc}
1\\&0&i\\&i&0\\&&&1
\end{array}\right)\!\!=
\exp\!\left[-i\mathcal H^{XY}_{i,j}\!(E_{i,j}^{XY}\!)
\frac{\pi}{E_{i,j}^{XY}}\right]\!. 
$$
It has been noted before that this gate is useful 
in order to generate more complex quantum 
operations~\cite{jens:PRL,kempe:enc_universality,echternach:cooperpairbox}.

By applying the \gate{iSWAP} gate twice, 
the \gate{CNOT} operation can be constructed
\begin{equation}
\label{eq:cnot_iswap}
\!\!\mbox{\placeimage{4em}{cnot.eps}}\!=\!\!\!\!
\mbox{\placeimage{4em}{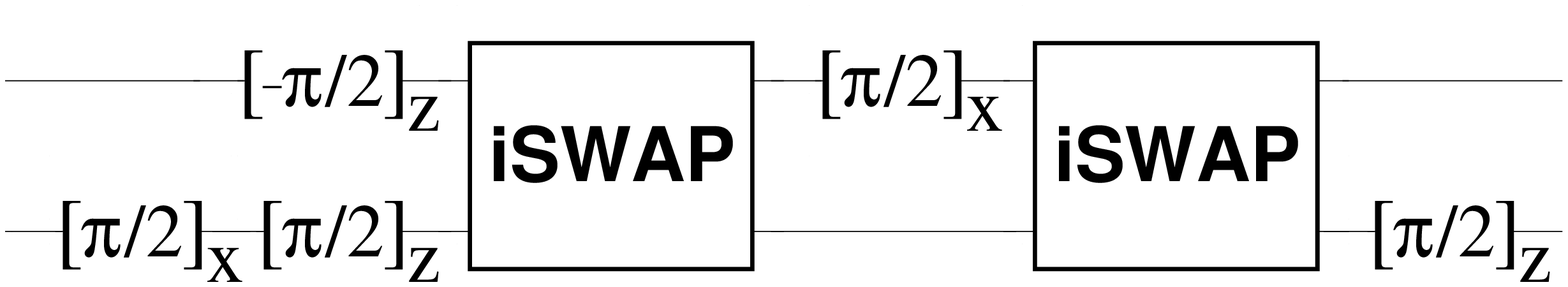}}
\end{equation}
We mention that in complex circuits the length of this sequence can be
reduced by noting that the ``outer'' one-bit operations partially
cancel out with preceding or subsequent one-bit gates.

Of course, also the \gate{SWAP} operation can be built with 
\gate{iSWAP} gates -- either by substituting Eq.~(\ref{eq:cnot_iswap})
in Eq.~(\ref{eq:swap_cnot}), or by applying 
the following sequence (which is considerably
shorter):
$$
\mbox{\placeimage{4em}{swap.eps}}=
$$
\vspace*{-2em}
\begin{equation}
\label{eq:swap_w_iswap}
\mbox{\placeimage{4em}{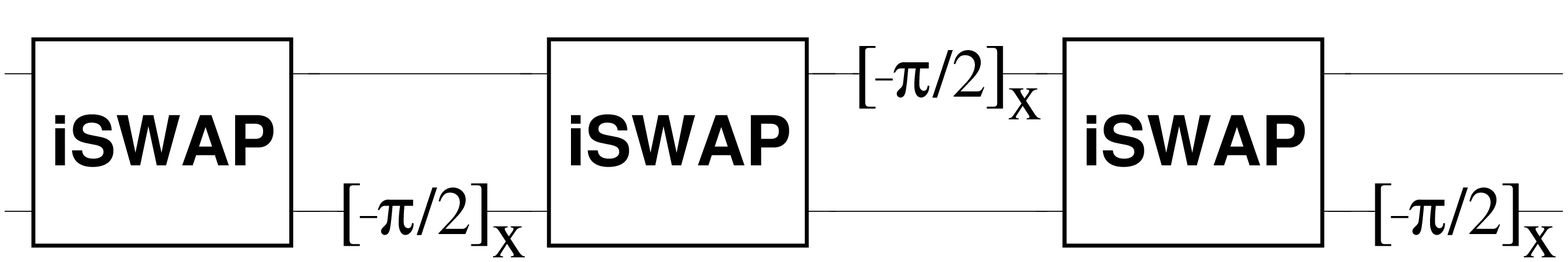}}\,.
\end{equation}

We mention that there exists also a proposal how to build the 
\gate{CNOT} gate by using $\sqrt{\mbox{\gate{iSWAP}}}$ 
(see Ref.~\cite{imamoglu:pap_w_sqrtiswap}).  

The close relation between the three types of Hamiltonians and the
corresponding two-bit operations has been demonstrated rigorously in
Ref.~\cite{vidal:interaction_cost}.

\section{\label{sec:cns_gate}
A natural gate for the $XY$ interaction}

In the following we focus on the $XY$ interaction.
As we have seen it can be used to generate a two-qubit gate
which (together with single-bit rotations) is sufficient for
universal quantum computation.
However,
neither \gate{CNOT} nor \gate{SWAP} can be realized
directly by using only the interaction part of the Hamiltonian, 
in contrast to
the $ZZ$ and the Heisenberg interaction.  
Now we are asking whether there exists a ``natural'' 
two-qubit operation similar to \gate{CNOT} also for the
$XY$-interaction, i.e.\ a gate which can be viewed as the quantum case of a
classical reversible operation like the \gate{CNOT} or the \gate{SWAP}
gate.

By analyzing the matrix of the \gate{iSWAP} gate we see that it can
be decomposed as 
$$
\mbox{\gate{iSWAP}}=
\left(\begin{array}{cccc}
    1\\&0&i\\&i&0\\&&&1
\end{array}\right)=
\left(\begin{array}{cccc}
    1\\&i\\&&i\\&&&1
\end{array}\right)\cdot
\left(\begin{array}{cccc}
    1\\&0&1\\&1&0\\&&&1
\end{array}\right).
$$
The second matrix represents the \gate{SWAP} operation while the 
first matrix is equivalent to \gate{CNOT} up to one-bit operations
since 
$$
\mbox{\placeimage{5em}{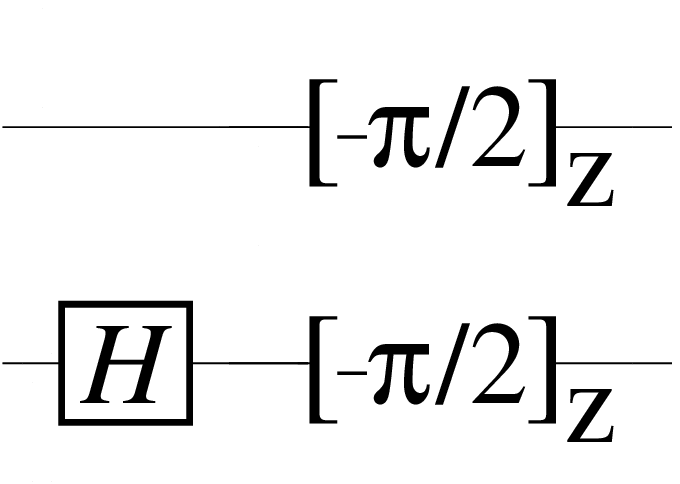}}\!\!\!
\left(\begin{array}{cccc}
			1&&&\\&i&&\\&&i&\\&&&1
	     \end{array}\right)\!\!\!\!\!\!
\mbox{\placeimage{5em}{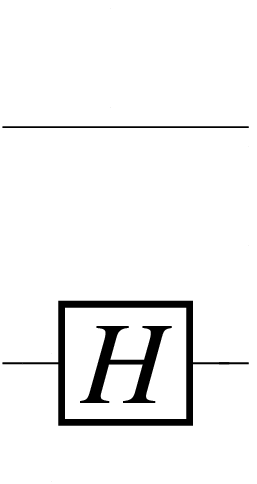}}
=
\mbox{\placeimage{5em}{cnot.eps}}\quad.
$$

Thus it follows that the \gate{iSWAP} gate is equivalent
to a combination of \gate{CNOT} and \gate{SWAP}. The exact 
sequence is
$$
\mbox{\gate{CNS}}:=\!\!\!
\mbox{\placeimage{5em}{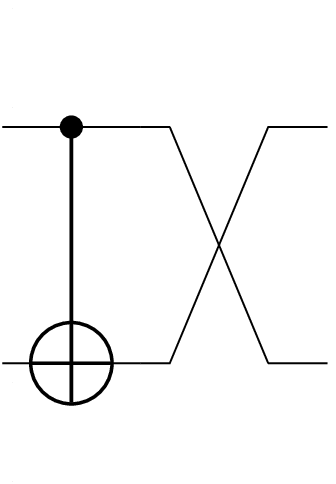}}
=\!\!\!
\mbox{\placeimage{5em}{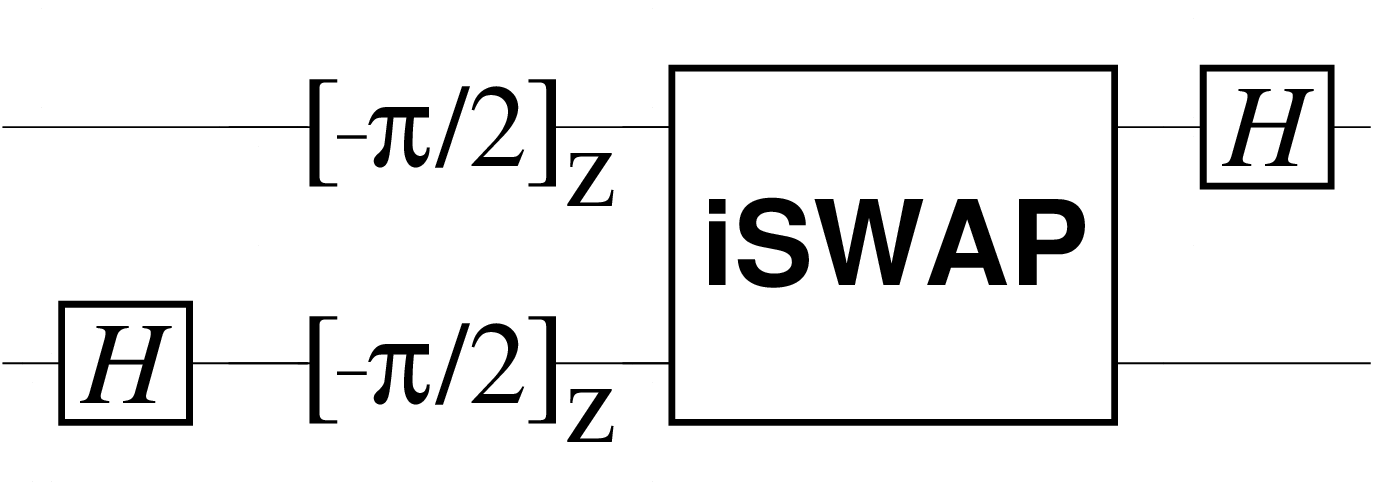}}
$$
For the sake of brevity, we introduce the name \gate{CNS} 
(``\gate{CNOT+SWAP}'')
for the new gate.

Remarkably this combined gate requires only a single operation
using the coupling Hamiltonian and can therefore be regarded
as a natural gate in the sense explained above.

One is tempted to object that the combination of \gate{CNOT} and 
\gate{SWAP} makes the \gate{CNS} gate  difficult to handle.
While this is true in principle, one may notice that in the
case of qubit couplings only between nearest neighbors it is
necessary to swap the qubit states anyhow 
(as we have discussed in Section~\ref{sec:introduction}).
Therefore one can try to exploit this feature by rearranging
the circuit in such a way that \gate{CNOT} and \gate{SWAP}
operations appear together and can be replaced by a
\gate{CNS} operation. 
Moreover, it should be mentioned that the
\gate{CNS} operation is considerably shorter than both the \gate{CNOT} and
the \gate{SWAP} operation (realized with the $XY$ coupling); 
so even with an overhead of two-bit operations
compared to a ``standard'' circuit (which uses \gate{CNOT} and \gate{SWAP})
the rearrangement of the network
may yield an advantage
in terms of the operation time required for the whole sequence.

\section{\label{sec:examples}Examples}

In this section we discuss two examples in order to demonstrate
that the \gate{CNS} gate derived above  is surprisingly powerful
in efficiently implementing quantum circuits in systems with
nearest-neighbor $XY$ interactions. 
As a simple example we first discuss the Toffoli gate.
In order to show that the method works  for more complex
networks as well, we then present an implementation of the five-bit
error correction found by DiVincenzo and Shor~\cite{divincenzo:5bit_ecc}.

\subsection{\label{subsec:toffoli}
    The three-bit Toffoli gate}

The three-bit Toffoli gate is the generalization of the
\gate{CNOT} gate with two control bits: it inverts the third bit if and only
if the first two bits are in the $|1\rangle$ state. It is of special interest
since it is the elementary gate for \emph{classical} reversible computation.
For this reason  it often appears in
circuits for tasks that also can be solved by classical reversible
computers, e.g.\ the modular exponentiation used in Shor's factoring 
algorithm~\cite{ekert:mod_exp_circ}. 

There are various proposals to implement the Toffoli gate. The shortest
one using the \gate{CNOT} gate as the only two-bit gate is
given in Ref.~\cite{divincenzo:q_gates_a_circ} 
(which is a simplification of the version 
in Ref.~\cite{barenco:univ_gates_f_qc}) 
and involves six \gate{CNOT} gates (the one-bit gates are given in 
Appendix~\ref{app:gates_for_toffoli}):
\begin{equation}
\label{eq:toffoli_w_cnot}
\mbox{\placeimage{6em}{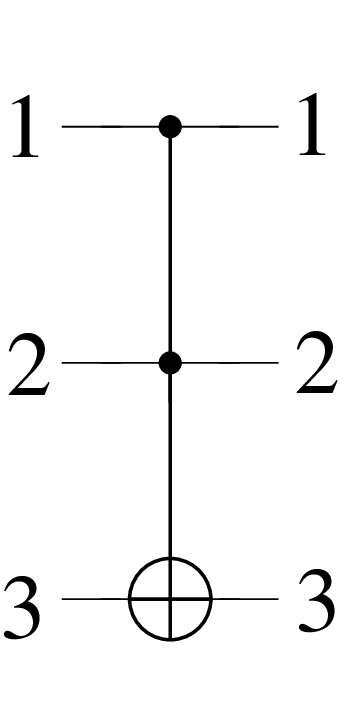}}
=
\mbox{\placeimage{6em}{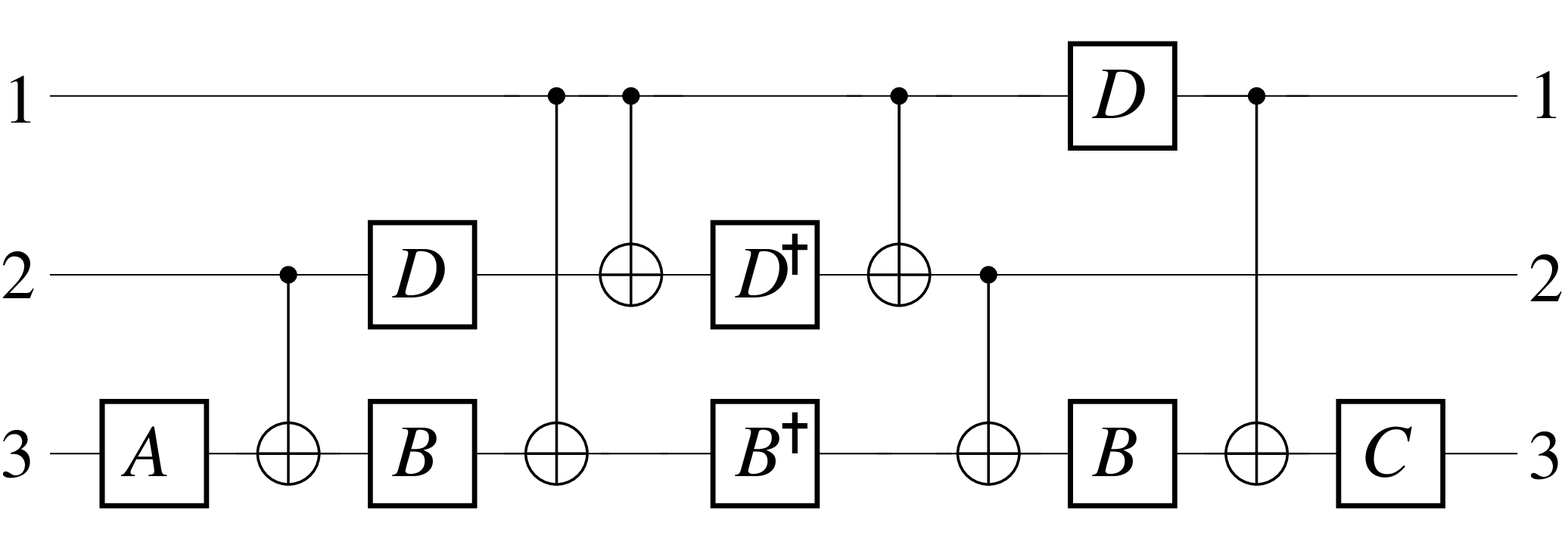}}
\end{equation}

We are considering systems with only nearest-neighbor
interaction. 
As the Toffoli gate often appears as an element 
in a circuit with more than three qubits, 
interaction will be possible only between the
qubit pairs 1-2 and 2-3, but not between qubits 1 and 3 (see
Eq.~(\ref{eq:toffoli_w_cnot})). 
Since in the circuit given above two of the \gate{CNOT} gates are acting
between qubits 1 and 3, one will have to swap one of the two qubits
with qubit 2 in order to make qubits 1 and 3 nearest neighbors. This
swapping  has to be undone after the \gate{CNOT} operation
in order to bring the qubits back in the right order. 
Therefore one has to
perform four \gate{SWAP} operations in addition to the six \gate{CNOT}s.
The number of \gate{CNOT} or \gate{iSWAP} gates to build a \gate{SWAP} gate
is three.  It turns out, however, that the \gate{CNOT} operations between
qubits $1$ and $3$ can be generated with only five (instead of seven)
nearest-neighbor \gate{CNOT}s. Thus one ends
up with 
\begin{itemize}
\item $4+2\cdot5=\mathbf{14}$ \gate{CNOT} gates or
\item $6\cdot2+4\cdot3=\mathbf{24}$ \gate{iSWAP} gates
\end{itemize}
required to obtain one Toffoli gate.

Now we rearrange the circuit to make use of the properties of
the \gate{CNS} gate.  
We have found the  sequence
\begin{equation}
\label{eq:toffoli_swap12}
\mbox{\placeimage{6em}{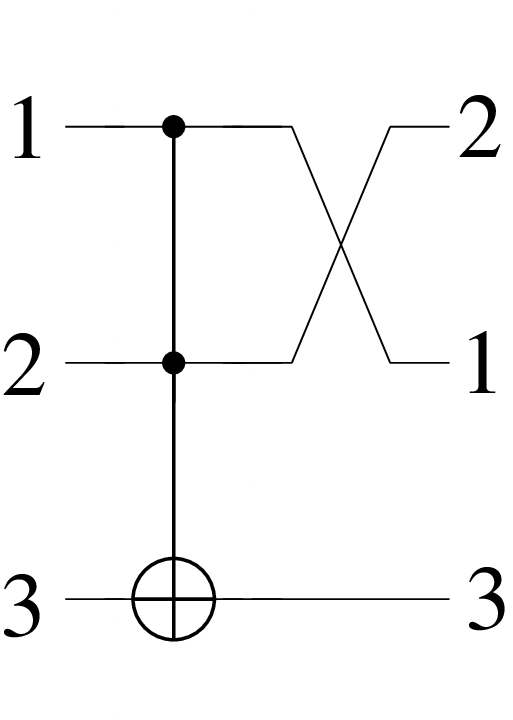}}
=
\end{equation}
\vspace{-1em}
$$
\!\!\!\!\!\mbox{\placeimage{6em}{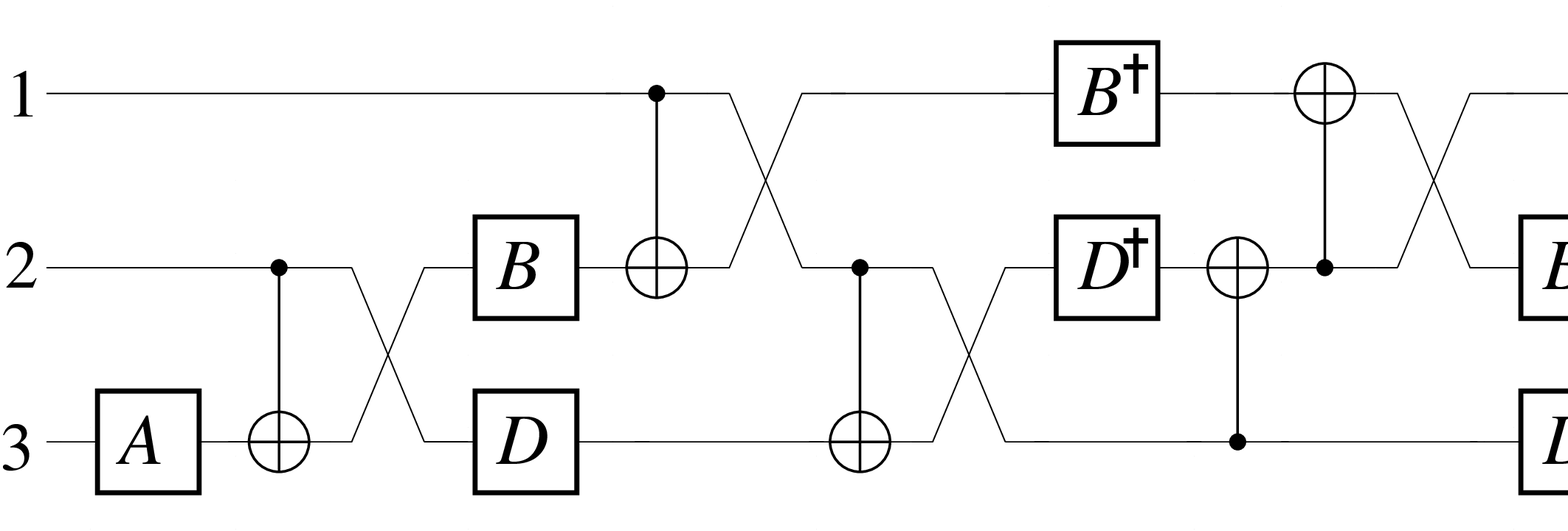}}
$$
which replaces five \gate{CNOT} operations by \gate{CNS} operations
and requires only a single additional \gate{SWAP} operation 
(i.e., in addition to the sequence in Eq.~(\ref{eq:toffoli_swap12}))
to correct for the fact that the sequence does not exactly 
implement the Toffoli gate but rather exchanges the qubits 1 and 2
(as can be seen easily by retracing the three lines
in the circuit above).
For the separate \gate{SWAP} three \gate{iSWAP}s have to be done
according to Eq.~(\ref{eq:swap_w_iswap}).
Thus, one finds that the total number of \gate{iSWAP} gates needed to implement
the Toffoli gate in a chain of qubits is
\begin{itemize}
\item $5\cdot1+1\cdot2+1\cdot3=\mathbf{10}$ \gate{iSWAP} gates
\end{itemize}
compared to the $24$ \gate{iSWAP} gates required for the ``naive version''
above.

Note that it is not necessarily a disadvantage to swap qubits 1 and 2
as in Eq.~(\ref{eq:toffoli_swap12}). Firstly, it can be regarded
as a combination of the Toffoli gate and \gate{SWAP} analogously
to \gate{CNS}.
In fact, by 
replacing some of the \gate{CNOT}s by \gate{CNS} gates  in the Toffoli 
network in  Eq.~(\ref{eq:toffoli_w_cnot}) 
every possible permutation of the input bits can be achieved 
(amusingly, \emph{except} the constant permutation), so one could try to
exploit this in more complex circuits the same way as the fact that \gate{CNS}
is a combination of \gate{CNOT} and \gate{SWAP}.

Secondly, in quantum computing there are
numerous circuits which first execute a sequence and then repeat
the operations in the reverse order, e.g., to reset some ancilla qubits. 
In this case swapping of two qubits is often irrelevant
(for examples of such circuits, see
the network for the $N$-bit Toffoli gate in 
Ref.~\cite{barenco:univ_gates_f_qc}
or the circuit of the quantum adder in Ref.~\cite{ekert:mod_exp_circ}).

\subsection{\label{subsec:5bit_ecc}
    A five-bit error-correcting code}

\begin{figure}
\includegraphics[width=25em]{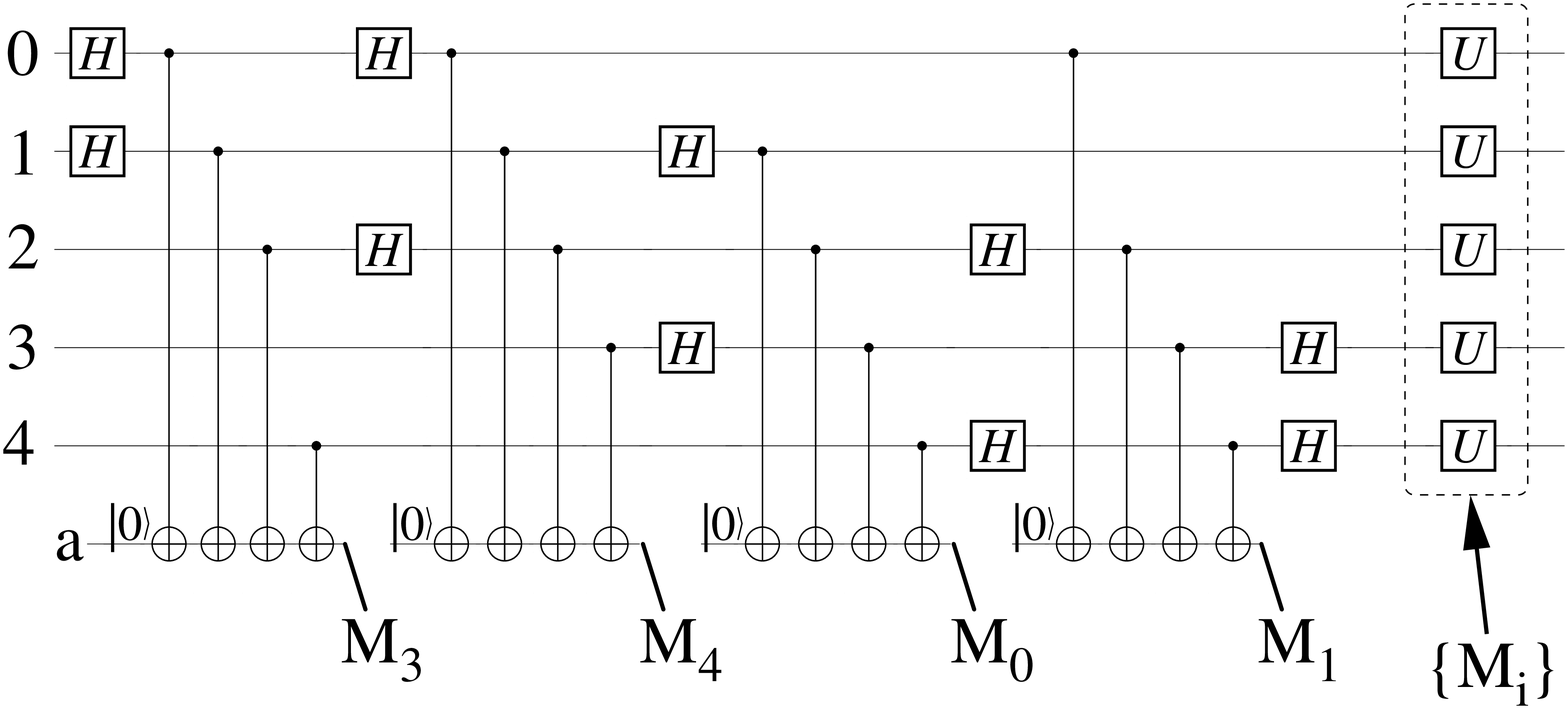}
\caption{\label{fig:5bit_ecc_network}
The five-bit error correction network as suggested by DiVincenzo and
Shor in~\cite{divincenzo:5bit_ecc}. The protected qubit state is encoded
in the qubits 0 to 4, and the measurements $M_i$ of the ancilla yield the
error syndrome which determines the correction operations $U$.
Instead of a single ancilla it is equally possible to use four different
ancillae such that each measurement is performed on a (physically) different
ancilla.}
\end{figure}

As a more advanced application where the 
use of \gate{CNS} gates gives a 
considerable advantage 
over the simple ``translation'' of the network, we
present an implementation of the five-bit error-correction network which
was found by DiVincenzo and Shor~\cite{divincenzo:5bit_ecc}. 
This network (see Fig.~\ref{fig:5bit_ecc_network})
can compensate
arbitrary one-bit errors as long as they occur only in one 
of the encoding qubits at a time.

The protected qubit is encoded in five physical qubits 0 to 4 as a 
superposition of five-qubit states. The error correction network 
makes use of four ancilla bits which are initialized to $|0\rangle$ before the
network is applied.  After carrying out the sequence of operations
the ancillae are measured (in the standard basis).

While the implementation of this circuit appears rather hopeless if there is
no direct interaction between the ancillae and \emph{each}  encoding qubit,
we will show that the \gate{CNS} gate
makes a straightforward implementation of this network possible. To this end, we
consider a setup of nine qubits (five qubits encoding the protected
state plus four ancilla bits) arranged in a ring, i.e.\
we have nearest-neighbor couplings with periodic boundary conditions. 

\begin{figure}
\placeimage{20em}{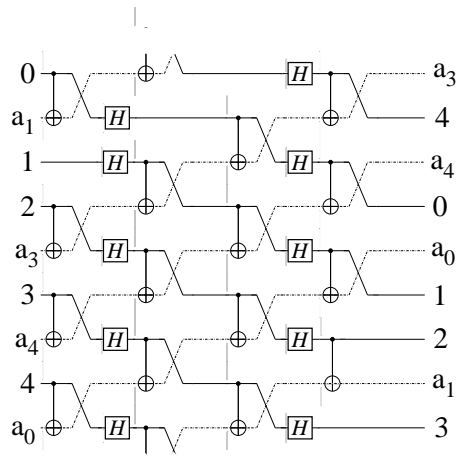}
\caption{\label{fig:5bit_ecc_cns}
Implementation of the five-bit error correcting code with \gate{CNS}
gates (see text).}
\end{figure}

By properly rearranging the gates, we obtain the circuit shown in 
Fig.~\ref{fig:5bit_ecc_cns}. 
The labels 0 to 4 on the lines denote the five qubits of the 
error-correcting code and, correspondingly, a$_0$, a$_1$, a$_3$ and a$_4$
 are the labels of the four ancilla bits for the  $M_i$s (see 
Fig.~\ref{fig:5bit_ecc_network}).
After the application of the network the four ancillae have to be measured
and the corresponding operations on the five bits to correct the error
syndrome have to be applied. Then the ancillae have to be reset to the 
$|0\rangle$ state.
For the whole setup, only one separate \gate{CNOT} is necessary, all other
\gate{CNOT} gates can be replaced by \gate{CNS} gates. 
Separate \gate{SWAP} operations
are not required at all. One may notice that
the two-bit operations in this network can be done in parallel which
makes the execution of the whole sequence considerably faster~%
\footnote{
Formally a parallel execution looks feasible also if the network can
be implemented with physical couplings between arbitrary pairs of
bits, in particular between the ancillae and \emph{each} encoding qubit.
Note, however, that in such schemes there is typically only \emph{one}
channel which mediates the coupling between the various qubits~%
\cite{imamoglu:pap_w_sqrtiswap,makhlin:jqubits_nature}. Consequently,
simultaneous execution of several two-qubit operations 
would result in $N$-qubit dynamics ($N>2$) which is to be excluded.
Therefore, in certain cases application of nearest-neighbor coupling appears 
to be even more powerful than coupling between arbitrary pairs of qubits.
}.

The equivalence of the networks in Fig.~\ref{fig:5bit_ecc_network}
and Fig.~\ref{fig:5bit_ecc_cns} is explained in detail 
in Appendix~\ref{app:rearr}.
Note that this implementation of the error-correcting 
code leaves the qubits in the original order, but \emph{rotates}
them by three bits. Therefore one has to keep track of the position of each bit.
In any case, after three subsequent applications of the error correction scheme
the encoding bits are back in their original order.
As to the ancilla bits, their order is changed in a well-defined
way. This has to be taken into account for the interpretation of the measurement
outcome. After the measurement, the order of the ancillae is 
irrelevant since they are re-initialized to $|0\rangle$. 

\begin{figure}
\includegraphics[width=25em]{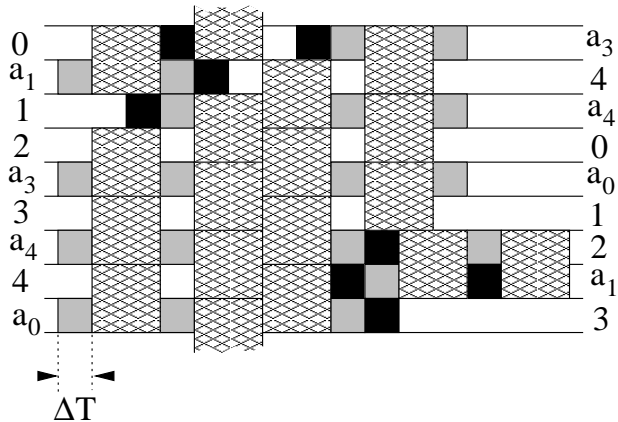}
\caption{\label{fig:5bit_ecc_hwseq}
Sequence of operations for the five-bit error correcting
code in Ref.~\cite{divincenzo:5bit_ecc} for a Hamiltonian with controllable
$\sigma_x$ and $\sigma_z$ part and nearest-neighbor $XY$ interaction 
between the qubits which are
arranged in a ring. The horizontal direction in the figure shows the
time axis while in the vertical direction the various actions on each qubit
are displayed. The boxes represent an active single-qubit or two-qubit
gate, while white spaces denote idle periods. Black boxes correspond to a 
$\sigma_z$ rotation, and gray boxes to a $\sigma_x$ rotation. 
The big hatched
boxes correspond to 
the action of an $XY$ coupling term between  two qubits.
The time grid is $\Delta T=0.5\,\pi\hbar/E_\mathrm{typ}$ where we
have assumed
equal energy scales for one-bit and two-bit Hamiltonians:
\protect $E_\mathrm{typ}^\mathrm{1-bit}\simeq 
          E_\mathrm{typ}^\mathrm{2-bit}\simeq E_\mathrm{typ}$. 
In practice one often finds
\protect $E_\mathrm{typ}^\mathrm{1-bit}\gg
          E_\mathrm{typ}^\mathrm{2-bit}$; in this case the width
	  of the one-bit blocks becomes vanishingly small.
The signs of the
corresponding energies are not contained in the diagram.
Two-bit blocks (representing \gate{iSWAP} gates) correspond to 
\gate{CNS} operations; the 
double \gate{iSWAP} block between  
qubits 2 and a$_1$ at the end of the 
sequence represents the separate \gate{CNOT} gate.
}
\end{figure}

Until now, we have achieved an efficient implementation of the 
five-bit error-correction
code  using the \gate{CNS} gate as an elementary building block. 
Let us conclude this discussion by studying the implementation
closer to the hardware level.
In order to realize the network in a setup of qubits with 
nearest-neighbor $XY$ interaction, one
can rewrite the circuit in Fig.~\ref{fig:5bit_ecc_cns} 
in terms of \gate{iSWAP} operations. 
The sequence can be simplified considerably (see 
Appendix~\ref{app:simplifications}). The result is illustrated
in Fig.~\ref{fig:5bit_ecc_hwseq}. 
If we assume that the energies in the one-bit and two-bit part
of the Hamiltionan are of the order $E_\mathrm{typ}^\mathrm{1-bit}$ and
$E_\mathrm{typ}^\mathrm{2-bit}$, respectively,
the total operation time of
the sequence (without measurement, correction step and resetting  the
ancillae) is $2.5\,\pi\hbar/E_\mathrm{typ}^\mathrm{1-bit}+
5\,\pi\hbar/E_\mathrm{typ}^\mathrm{2-bit}$. 

We mention that efficient solutions for similar tasks in error correction
have been developed also in Refs.~\cite{braunstein:24pulses,
burkard:opt_3bit_ecc}.

\section{Conclusions}

A considerable part of the research efforts in quantum computation
is devoted either to the development of quantum circuits to formally
solve a given computational task (one could call this sector `development
of algorithms'), or to the practical implementation
of  specific algorithms for a given type of hardware.
We have studied questions residing in the domain of problems
between these two sectors. We have focused on aspects 
which are common to different types of
hardware and which may affect the feasibility of the practical
implementation of  existing formal solutions.

In particular we have discussed two hardware-related problems which 
seem to render the physical implementation of  quantum networks difficult: 
\emph{i)} coupling between arbitrary pairs of qubits cannot be achieved for
all hardware proposals; \emph{ii)} depending on the implementation, certain
two-qubit gates are more or less difficult to realize.

While at first glance these hardware properties seem
to imply that hardware for quantum computation has to meet very
specific demands that are hard to realize, our results indicate 
a much more optimistic conclusion. We have demonstrated
that  the $XY$ interaction---which is neither capable 
of realizing a \gate{CNOT} nor a \gate{SWAP} gate directly---is 
surprisingly powerful in implementing certain quantum networks.
This has been achieved by realizing that the \gate{CNS} gate
is a combination of the \gate{CNOT} gate and the \gate{SWAP}
operation which can be obtained with a single two-qubit step.

Moreover, our examples illustrate that this gate overcomes the
problems arising from the hardware property \emph{i)} mentioned
above in a natural way. 
The built-in \gate{SWAP} makes an 
efficient implementation possible even if  only nearest-neighbor 
interaction between the qubits is available. 
This approach may devise ways how
to tackle this difficulty also for other types of two-qubit interactions.
Interestingly, 
our solutions display the possibility of parallel execution of operations 
at the hardware level.
Finally we mention that the same methods work equally well also
for networks other than 
those presented in this article. Similar solutions can be found, e.g., for
the Quantum Fourier Transform 
(see, e.g., Ref.~\cite{cleve:q_alg_revisited}),
for the quantum adder described in 
Ref.~\cite{draper:qc_adder} 
(which is adding two quantum numbers), 
or for an adder which is adding one classical to one quantum 
number~\cite{beauregard:shor_2n3}.
This suggests
the applicability of the ideas presented here for a wide range
of quantum networks.

\section*{Acknowledgement}

We are grateful to J.~L.~Dodd for pointing out to us Ref.~%
\cite{brylinski:universal_gates}.

\appendix

\section{One-bit gates used for the Toffoli gate
    \label{app:gates_for_toffoli}}

The one-bit gates which were used for the three-bit
Toffoli gate in Section~\ref{subsec:toffoli}
are (up to normalization factors):
$$
\begin{array}{cc}
A=\left(\begin{array}{cc}1\\&i\end{array}\right)\ , 
     &
B=\left(\begin{array}{cc}1&1-\sqrt2\\\sqrt2-1&1\end{array}\right)\ ,
     \\&\\
C=\left(\begin{array}{cc}1&\sqrt2-1\\i(\sqrt2-1)&-i\end{array}\right)\ ,
     &
D=\left(\begin{array}{cc}1&0\\0&e^{-i\pi/4}\end{array}\right)\ .
\end{array}
$$

\section{\label{app:rearr}Equivalence of the error correcting networks}

We need to show that the two error correction
circuits in 
Fig.~\ref{fig:5bit_ecc_network} and Fig.~\ref{fig:5bit_ecc_cns} 
are equivalent, i.e., that they correspond to the same unitary. 
To this end we note that  gates which have no bits in 
common commute trivially. Further, 
two subsequent \gate{CNOT} operations commute if they
act on the same target
bit (in our case this means that they have the ancilla 
in common). 
Also two \gate{CNOT} gates with the same control
bit and different targets commute as long as the control bit is
not modified by a {\em single} Hadamard operation  between them.

Starting from these observations, the question whether
the two error correcting networks are identical reduces 
to proving that the two marked blocks in
\begin{equation}
\placeimage{7em}{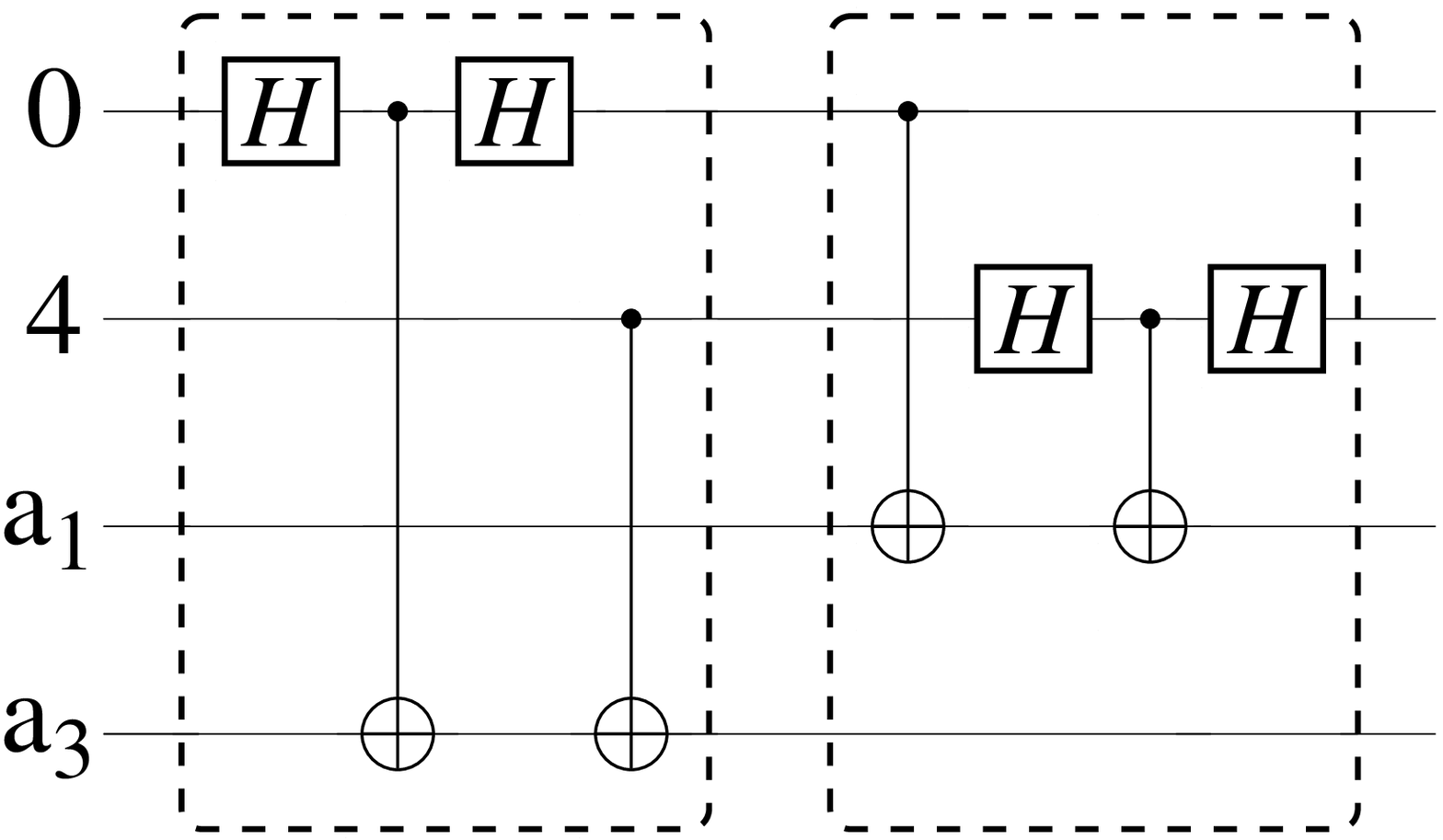}
\label{eq:app_c_commreq}
\end{equation}
do commute. 

This can be seen as follows. First, choose the
Hadamard transformed basis for the ancillae (we will denote them by
$\hat a_i$). In this basis, the \gate{CNOT} gates originally enclosed
by Hadamard operations  become \gate{CNOT}s with control 
and target reversed, i.e., controlled $\sigma_x$ operations.
The other \gate{CNOT} gates turn into controlled phase
flips (i.e., controlled $\sigma_z$ operations), analogously. 
Thus, the network in Eq.~(\ref{eq:app_c_commreq}) corresponds to
\begin{equation}
\placeimage{7em}{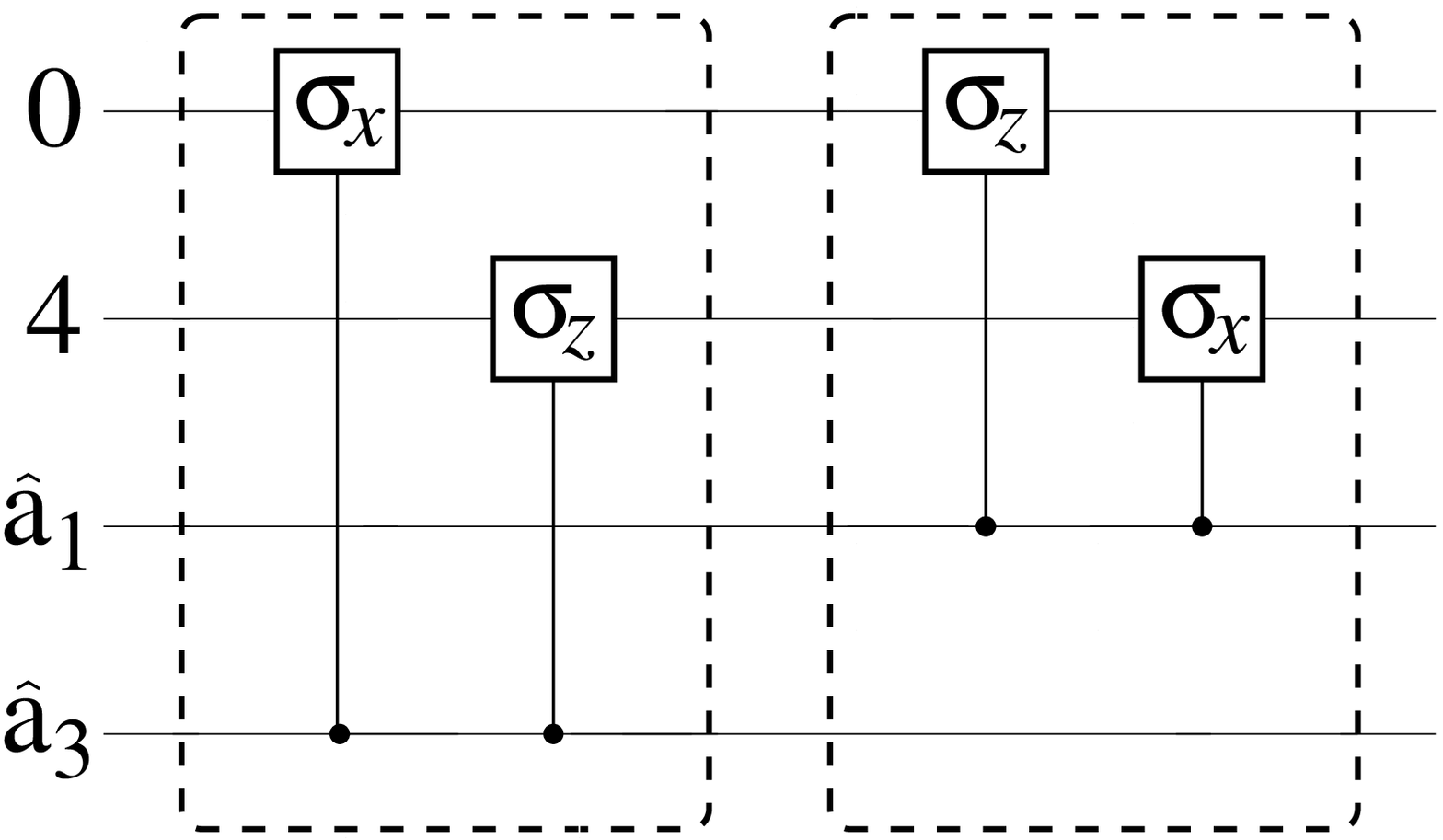}\ .
\label{eq:app_c_commreq2}
\end{equation}
In this representation, it becomes obvious that the two
blocks indeed do commute:
if at least one $\hat
a_i$ is zero, one of the two controlled operations on each qubit
$0$ and $4$ is the identity, and the operations commute. 
On the other hand, if $\hat a_1=\hat a_3=1$, 
exchanging $\sigma_x$ and
$\sigma_z$ on {\it one} bit results in a global minus sign. 
Changing the order of the two blocks in 
Eq.~(\ref{eq:app_c_commreq2}) corresponds to two 
simultaneous changes of this kind and 
leaves the all-over result
unchanged.  Therefore, 
the networks in Fig.~\ref{fig:5bit_ecc_network} and 
Fig.~\ref{fig:5bit_ecc_cns} are identical.

\section{Simplifications for circuits with \gate{iSWAP}
    \label{app:simplifications}}

In this appendix, we briefly describe the methods which
can be used for the simplification of a network like the one presented for
the five-bit error correction circuit in Section~\ref{subsec:5bit_ecc}.
We will consider rotations about the $x$ and the $z$-axis,
and the \gate{iSWAP} operation as the basic 
building blocks for the implementation. 
All other operations will be expressed in
terms of these operations. Although the
application of the ideas is rather straightforward it 
is difficult to provide formal recipes how to use them.

\emph{1. One-bit simplifications.\quad} 
First, rotations about the same axis can be collected, e.g.:
$$
\mbox{\placeimage{1.5em}{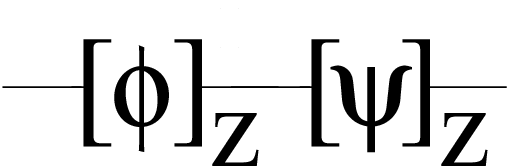}}=
\mbox{\placeimage{1.5em}{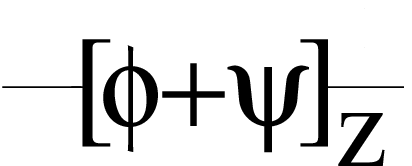}}
$$
Clearly, rotations 
by an angle of $2\pi$ can be dropped. Operation time can be saved
by applying a $-\pi/2$ rotation instead of a  $+3\pi/2$ one. 

A more sophisticated problem is the simplification of compound expressions
of $x$ \emph{and} $z$-rotations. To this end, the following 
ways to express the Hadamard transformation are useful:
$$
\!\!\!\!\mbox{\placeimage{1.5em}{hadamard.eps}}\!\!=\!\!\!\!\!\!
\mbox{\placeimage{1.5em}{had_equiv.eps}}\!\!=\!\!\!\!\!\!
\mbox{\placeimage{1.5em}{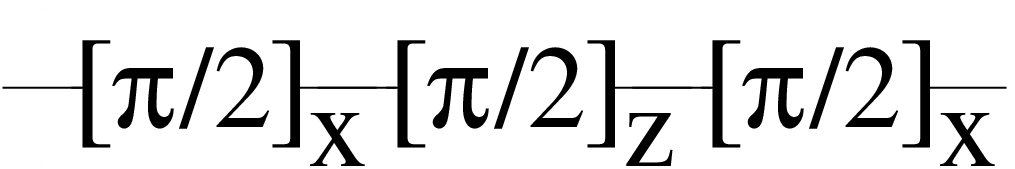}}\!\!=
$$
\vspace*{-1.5em}
$$
\quad\ =\!\!\!\!\!\!
\mbox{\placeimage{1.5em}{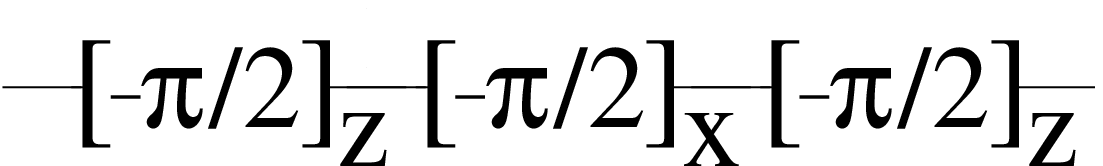}}\!\!=\!\!\!\!\!\!
\mbox{\placeimage{1.5em}{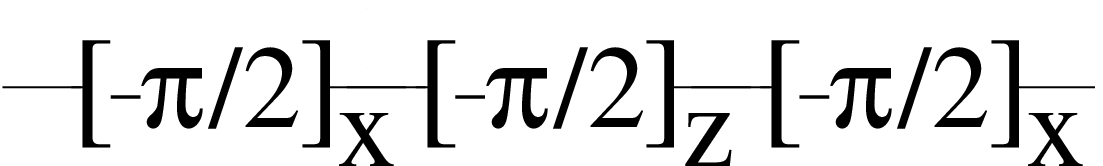}}
$$
By choosing the appropriate way to represent the Hadamard transformations in
the network (or by sometimes ``artificially'' creating one of these triples, 
for example by inserting a pair of operations whose product is unity)
and replacing it by another one,
considerable simplifications can be achieved. 
The applicability of these simplifications becomes particularly apparent if
the single-bit operations are considered in the context of two-bit operations.

\emph{2. Two-bit simplifications.\quad} There is essentially only one way to 
simplify two-bit  expressions for circuits containing \gate{iSWAP}:
$$
\mbox{\placeimage{4em}{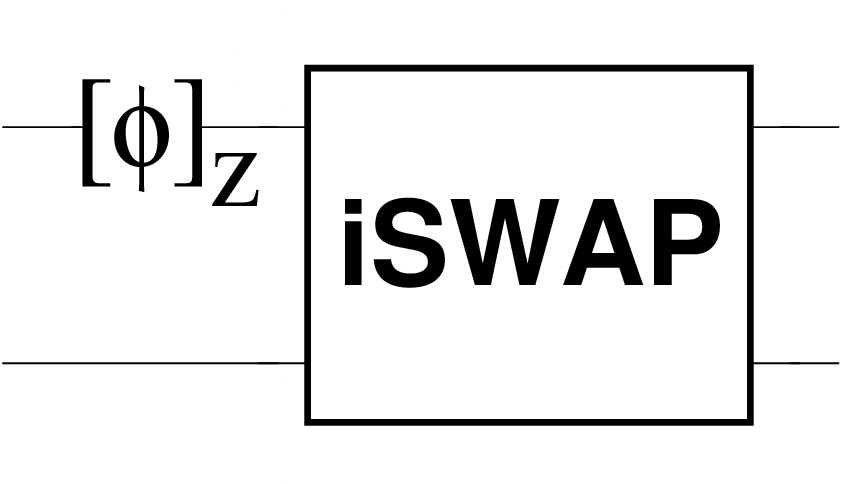}}\,\,\,=
\mbox{\placeimage{4em}{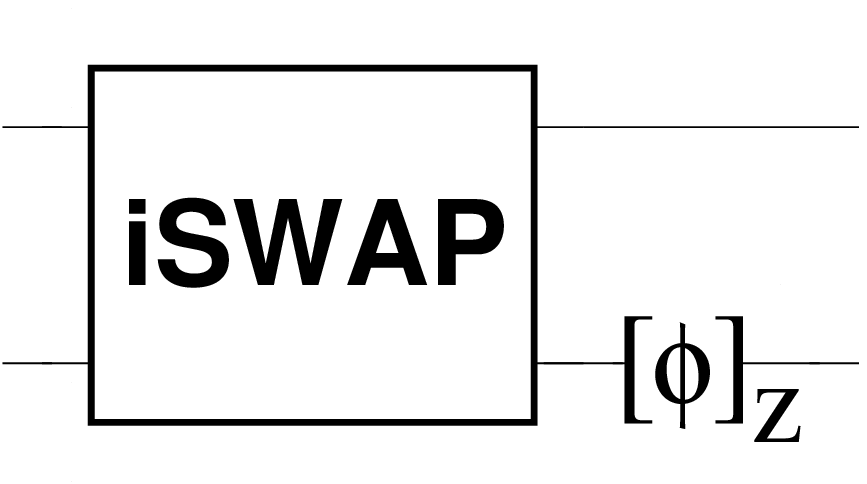}}\quad,
$$
i.e.,  $z$-rotations ``commute'' with \gate{iSWAP} if  simultaneosly
the $z$-rotation is flipped to the other qubit.

\emph{3. Ancilla simplifications.\quad}
Finally, one can apply also
one-bit simplifications to the ancilla bits which are possible due to
the fact that we know the initial state of the ancilla and, moreover, 
the ancillae are measured in the $\{|0\rangle,|1\rangle\}$ basis
at the end.

At the beginning of the error-correction sequence, the ancillae are 
set to $|0\rangle$. Therefore, $z$-rotations immediately 
after the initialization can be omitted (since
global phases are not important). 

Further, let us assume the ancilla is in the state $a|0\rangle+b|1\rangle$
just before the measurement. As the measurement is
performed in the $\{|0\rangle,|1\rangle\}$ basis, a  $z$ rotation
before the measurement would not affect the result. Therefore,
also these $z$ rotation need not be considered.

\bibliography{paper}

\begin{thebibliography}{25}
\expandafter\ifx\csname natexlab\endcsname\relax\def\natexlab#1{#1}\fi
\expandafter\ifx\csname bibnamefont\endcsname\relax
  \def\bibnamefont#1{#1}\fi
\expandafter\ifx\csname bibfnamefont\endcsname\relax
  \def\bibfnamefont#1{#1}\fi
\expandafter\ifx\csname citenamefont\endcsname\relax
  \def\citenamefont#1{#1}\fi
\expandafter\ifx\csname url\endcsname\relax
  \def\url#1{\texttt{#1}}\fi
\expandafter\ifx\csname urlprefix\endcsname\relax\def\urlprefix{URL }\fi
\providecommand{\bibinfo}[2]{#2}
\providecommand{\eprint}[2][]{\url{#2}}

\bibitem[{\citenamefont{Brylinski and
  Brylinski}(2002)}]{brylinski:universal_gates}
\bibinfo{author}{\bibfnamefont{J.-L.} \bibnamefont{Brylinski}}
  \bibnamefont{and}
  \bibinfo{author}{\bibfnamefont{R.}~\bibnamefont{Brylinski}}, in
  \emph{\bibinfo{booktitle}{Mathematics of Quantum Computation}}, edited by
  \bibinfo{editor}{\bibfnamefont{R.}~\bibnamefont{Brylinski}} \bibnamefont{and}
  \bibinfo{editor}{\bibfnamefont{G.}~\bibnamefont{Chen}}
  (\bibinfo{publisher}{Chapman and Hall/CRC Press}, \bibinfo{year}{2002}),
  \eprint{quant-ph/0108062}.

\bibitem[{\citenamefont{Bremner et~al.}(2002)\citenamefont{Bremner, Dawson,
  Dodd, Gilchrist, Harrow, Mortimer, Nielsen, and
  Osborne}}]{bremner:all_are_universal}
\bibinfo{author}{\bibfnamefont{M.~J.} \bibnamefont{Bremner}},
  \bibinfo{author}{\bibfnamefont{C.~M.} \bibnamefont{Dawson}},
  \bibinfo{author}{\bibfnamefont{J.~L.} \bibnamefont{Dodd}},
  \bibinfo{author}{\bibfnamefont{A.}~\bibnamefont{Gilchrist}},
  \bibinfo{author}{\bibfnamefont{A.~W.} \bibnamefont{Harrow}},
  \bibinfo{author}{\bibfnamefont{D.}~\bibnamefont{Mortimer}},
  \bibinfo{author}{\bibfnamefont{M.~A.} \bibnamefont{Nielsen}},
  \bibnamefont{and} \bibinfo{author}{\bibfnamefont{T.~J.}
  \bibnamefont{Osborne}}, \bibinfo{journal}{Phys. Rev. Lett.}
  \textbf{\bibinfo{volume}{89}}, \bibinfo{pages}{247902}
  (\bibinfo{year}{2002}), \eprint{quant-ph/0207072}.

\bibitem[{\citenamefont{Barenco et~al.}(1995)\citenamefont{Barenco, Bennett,
  Cleve, DiVincenzo, Margolus, Shor, Sleator, Smolin, and
  Weinfurter}}]{barenco:univ_gates_f_qc}
\bibinfo{author}{\bibfnamefont{A.}~\bibnamefont{Barenco}},
  \bibinfo{author}{\bibfnamefont{C.}~\bibnamefont{Bennett}},
  \bibinfo{author}{\bibfnamefont{R.}~\bibnamefont{Cleve}},
  \bibinfo{author}{\bibfnamefont{D.}~\bibnamefont{DiVincenzo}},
  \bibinfo{author}{\bibfnamefont{N.}~\bibnamefont{Margolus}},
  \bibinfo{author}{\bibfnamefont{P.}~\bibnamefont{Shor}},
  \bibinfo{author}{\bibfnamefont{T.}~\bibnamefont{Sleator}},
  \bibinfo{author}{\bibfnamefont{J.}~\bibnamefont{Smolin}}, \bibnamefont{and}
  \bibinfo{author}{\bibfnamefont{H.}~\bibnamefont{Weinfurter}},
  \bibinfo{journal}{Phys. Rev. A} \textbf{\bibinfo{volume}{52}},
  \bibinfo{pages}{3457} (\bibinfo{year}{1995}).

\bibitem[{\citenamefont{Loss and DiVincenzo}(1998)}]{loss:qc_w_qdots}
\bibinfo{author}{\bibfnamefont{D.}~\bibnamefont{Loss}} \bibnamefont{and}
  \bibinfo{author}{\bibfnamefont{D.}~\bibnamefont{DiVincenzo}},
  \bibinfo{journal}{Phys. Rev. A} \textbf{\bibinfo{volume}{57}},
  \bibinfo{pages}{120} (\bibinfo{year}{1998}).

\bibitem[{\citenamefont{Biolatti et~al.}(2000)\citenamefont{Biolatti, Iotti,
  Zanardi, and Rossi}}]{rossi:excitons}
\bibinfo{author}{\bibfnamefont{E.}~\bibnamefont{Biolatti}},
  \bibinfo{author}{\bibfnamefont{R.}~\bibnamefont{Iotti}},
  \bibinfo{author}{\bibfnamefont{P.}~\bibnamefont{Zanardi}}, \bibnamefont{and}
  \bibinfo{author}{\bibfnamefont{F.}~\bibnamefont{Rossi}},
  \bibinfo{journal}{Phys. Rev. Lett.} \textbf{\bibinfo{volume}{85}},
  \bibinfo{pages}{5647} (\bibinfo{year}{2000}).

\bibitem[{\citenamefont{Orlando et~al.}(1999)\citenamefont{Orlando, Mooij,
  Tian, van~der Wal, Levitov, Lloyd, and Mazo}}]{orlando:flux_qubits}
\bibinfo{author}{\bibfnamefont{T.~P.} \bibnamefont{Orlando}},
  \bibinfo{author}{\bibfnamefont{J.~E.} \bibnamefont{Mooij}},
  \bibinfo{author}{\bibfnamefont{L.}~\bibnamefont{Tian}},
  \bibinfo{author}{\bibfnamefont{C.~H.} \bibnamefont{van~der Wal}},
  \bibinfo{author}{\bibfnamefont{L.~S.} \bibnamefont{Levitov}},
  \bibinfo{author}{\bibfnamefont{S.}~\bibnamefont{Lloyd}}, \bibnamefont{and}
  \bibinfo{author}{\bibfnamefont{J.~J.} \bibnamefont{Mazo}},
  \bibinfo{journal}{Phys. Rev. B} \textbf{\bibinfo{volume}{60}},
  \bibinfo{pages}{15398} (\bibinfo{year}{1999}).

\bibitem[{\citenamefont{Siewert et~al.}(2000)\citenamefont{Siewert, Fazio,
  Palma, and Sciacca}}]{jens:JLTP}
\bibinfo{author}{\bibfnamefont{J.}~\bibnamefont{Siewert}},
  \bibinfo{author}{\bibfnamefont{R.}~\bibnamefont{Fazio}},
  \bibinfo{author}{\bibfnamefont{G.~M.} \bibnamefont{Palma}}, \bibnamefont{and}
  \bibinfo{author}{\bibfnamefont{E.}~\bibnamefont{Sciacca}},
  \bibinfo{journal}{J. Low. Temp. Phys.} \textbf{\bibinfo{volume}{118}},
  \bibinfo{pages}{795} (\bibinfo{year}{2000}).

\bibitem[{\citenamefont{Makhlin et~al.}(1999)\citenamefont{Makhlin, Sch{\"o}n,
  and Shnirman}}]{makhlin:jqubits_nature}
\bibinfo{author}{\bibfnamefont{Y.}~\bibnamefont{Makhlin}},
  \bibinfo{author}{\bibfnamefont{G.}~\bibnamefont{Sch{\"o}n}},
  \bibnamefont{and} \bibinfo{author}{\bibfnamefont{A.}~\bibnamefont{Shnirman}},
  \bibinfo{journal}{Nature} \textbf{\bibinfo{volume}{398}},
  \bibinfo{pages}{305} (\bibinfo{year}{1999}).

\bibitem[{\citenamefont{Kane}(1998)}]{kane:silic_nuclspin_qc}
\bibinfo{author}{\bibfnamefont{B.}~\bibnamefont{Kane}},
  \bibinfo{journal}{Nature} \textbf{\bibinfo{volume}{393}},
  \bibinfo{pages}{133} (\bibinfo{year}{1998}).

\bibitem[{\citenamefont{Vrijen et~al.}(2000)\citenamefont{Vrijen, Yablonovitch,
  Wang, Jiang, Balandin, Roychowdhury, Mor, and
  DiVincenzo}}]{vrijen:spin_res_trans}
\bibinfo{author}{\bibfnamefont{R.}~\bibnamefont{Vrijen}},
  \bibinfo{author}{\bibfnamefont{E.}~\bibnamefont{Yablonovitch}},
  \bibinfo{author}{\bibfnamefont{K.}~\bibnamefont{Wang}},
  \bibinfo{author}{\bibfnamefont{H.~W.} \bibnamefont{Jiang}},
  \bibinfo{author}{\bibfnamefont{A.}~\bibnamefont{Balandin}},
  \bibinfo{author}{\bibfnamefont{V.}~\bibnamefont{Roychowdhury}},
  \bibinfo{author}{\bibfnamefont{T.}~\bibnamefont{Mor}}, \bibnamefont{and}
  \bibinfo{author}{\bibfnamefont{D.}~\bibnamefont{DiVincenzo}},
  \bibinfo{journal}{Phys. Rev. A} \textbf{\bibinfo{volume}{62}},
  \bibinfo{pages}{012306} (\bibinfo{year}{2000}).

\bibitem[{\citenamefont{Imamo{\=g}lu et~al.}(1999)\citenamefont{Imamo{\=g}lu,
  Awschalom, Burkard, DiVincenzo, Loss, Sherwin, and
  Small}}]{imamoglu:pap_w_sqrtiswap}
\bibinfo{author}{\bibfnamefont{A.}~\bibnamefont{Imamo{\=g}lu}},
  \bibinfo{author}{\bibfnamefont{D.~D.} \bibnamefont{Awschalom}},
  \bibinfo{author}{\bibfnamefont{G.}~\bibnamefont{Burkard}},
  \bibinfo{author}{\bibfnamefont{D.~P.} \bibnamefont{DiVincenzo}},
  \bibinfo{author}{\bibfnamefont{D.}~\bibnamefont{Loss}},
  \bibinfo{author}{\bibfnamefont{M.}~\bibnamefont{Sherwin}}, \bibnamefont{and}
  \bibinfo{author}{\bibfnamefont{A.}~\bibnamefont{Small}},
  \bibinfo{journal}{Phys. Rev. Lett.} \textbf{\bibinfo{volume}{83}},
  \bibinfo{pages}{4204} (\bibinfo{year}{1999}).

\bibitem[{\citenamefont{Mozyrsky et~al.}(2001)\citenamefont{Mozyrsky, Privman,
  and Glasser}}]{mozyrsky:nspin_2deg}
\bibinfo{author}{\bibfnamefont{D.}~\bibnamefont{Mozyrsky}},
  \bibinfo{author}{\bibfnamefont{V.}~\bibnamefont{Privman}}, \bibnamefont{and}
  \bibinfo{author}{\bibfnamefont{M.}~\bibnamefont{Glasser}},
  \bibinfo{journal}{Phys. Rev. Lett.} \textbf{\bibinfo{volume}{86}},
  \bibinfo{pages}{5112} (\bibinfo{year}{2001}).

\bibitem[{\citenamefont{Makhlin}(2000)}]{makhlin:2bit_equiv}
\bibinfo{author}{\bibfnamefont{Y.}~\bibnamefont{Makhlin}}
  (\bibinfo{year}{2000}), \eprint{quant-ph/0002045}.

\bibitem[{\citenamefont{Siewert and Fazio}(2001)}]{jens:PRL}
\bibinfo{author}{\bibfnamefont{J.}~\bibnamefont{Siewert}} \bibnamefont{and}
  \bibinfo{author}{\bibfnamefont{R.}~\bibnamefont{Fazio}},
  \bibinfo{journal}{Phys. Rev. Lett.} \textbf{\bibinfo{volume}{87}},
  \bibinfo{pages}{257905} (\bibinfo{year}{2001}).

\bibitem[{\citenamefont{Kempe et~al.}(2001)\citenamefont{Kempe, Bacon,
  DiVincenzo, and Whaley}}]{kempe:enc_universality}
\bibinfo{author}{\bibfnamefont{J.}~\bibnamefont{Kempe}},
  \bibinfo{author}{\bibfnamefont{D.}~\bibnamefont{Bacon}},
  \bibinfo{author}{\bibfnamefont{D.~P.} \bibnamefont{DiVincenzo}},
  \bibnamefont{and} \bibinfo{author}{\bibfnamefont{K.}~\bibnamefont{Whaley}},
  \bibinfo{journal}{Quantum Inf.\ Comput.} \textbf{\bibinfo{volume}{1}},
  \bibinfo{pages}{33} (\bibinfo{year}{2001}), \eprint{quant-ph/0112013}.

\bibitem[{\citenamefont{Echternach et~al.}(2001)\citenamefont{Echternach,
  Williams, Dultz, Delsing, Braunstein, and
  Dowling}}]{echternach:cooperpairbox}
\bibinfo{author}{\bibfnamefont{P.}~\bibnamefont{Echternach}},
  \bibinfo{author}{\bibfnamefont{C.~P.} \bibnamefont{Williams}},
  \bibinfo{author}{\bibfnamefont{S.~C.} \bibnamefont{Dultz}},
  \bibinfo{author}{\bibfnamefont{P.}~\bibnamefont{Delsing}},
  \bibinfo{author}{\bibfnamefont{S.}~\bibnamefont{Braunstein}},
  \bibnamefont{and} \bibinfo{author}{\bibfnamefont{J.~P.}
  \bibnamefont{Dowling}}, \bibinfo{journal}{Quantum Inf.\ Comput.}
  \textbf{\bibinfo{volume}{1}}, \bibinfo{pages}{143} (\bibinfo{year}{2001}),
  \eprint{quant-ph/0112025}.

\bibitem[{\citenamefont{Vidal et~al.}(2002)\citenamefont{Vidal, Hammerer, and
  Cirac}}]{vidal:interaction_cost}
\bibinfo{author}{\bibfnamefont{G.}~\bibnamefont{Vidal}},
  \bibinfo{author}{\bibfnamefont{K.}~\bibnamefont{Hammerer}}, \bibnamefont{and}
  \bibinfo{author}{\bibfnamefont{J.~I.} \bibnamefont{Cirac}},
  \bibinfo{journal}{Phys. Rev. Lett.} \textbf{\bibinfo{volume}{88}},
  \bibinfo{pages}{237902} (\bibinfo{year}{2002}), \eprint{quant-ph/0112168}.

\bibitem[{\citenamefont{DiVincenzo and Shor}(1996)}]{divincenzo:5bit_ecc}
\bibinfo{author}{\bibfnamefont{D.}~\bibnamefont{DiVincenzo}} \bibnamefont{and}
  \bibinfo{author}{\bibfnamefont{P.}~\bibnamefont{Shor}},
  \bibinfo{journal}{Phys. Rev. Lett.} \textbf{\bibinfo{volume}{77}},
  \bibinfo{pages}{3260} (\bibinfo{year}{1996}).

\bibitem[{\citenamefont{Vedral et~al.}(1996)\citenamefont{Vedral, Barenco, and
  Ekert}}]{ekert:mod_exp_circ}
\bibinfo{author}{\bibfnamefont{V.}~\bibnamefont{Vedral}},
  \bibinfo{author}{\bibfnamefont{A.}~\bibnamefont{Barenco}}, \bibnamefont{and}
  \bibinfo{author}{\bibfnamefont{A.}~\bibnamefont{Ekert}},
  \bibinfo{journal}{Phys. Rev. A} \textbf{\bibinfo{volume}{54}},
  \bibinfo{pages}{147} (\bibinfo{year}{1996}).

\bibitem[{\citenamefont{DiVincenzo}(1998)}]{divincenzo:q_gates_a_circ}
\bibinfo{author}{\bibfnamefont{D.}~\bibnamefont{DiVincenzo}},
  \bibinfo{journal}{Proc.\ R.\ Soc.\ London A} \textbf{\bibinfo{volume}{454}},
  \bibinfo{pages}{261} (\bibinfo{year}{1998}).

\bibitem[{\citenamefont{Braunstein and Smolin}(1997)}]{braunstein:24pulses}
\bibinfo{author}{\bibfnamefont{S.~L.} \bibnamefont{Braunstein}}
  \bibnamefont{and} \bibinfo{author}{\bibfnamefont{J.~A.}
  \bibnamefont{Smolin}}, \bibinfo{journal}{Phys. Rev. A}
  \textbf{\bibinfo{volume}{55}}, \bibinfo{pages}{945} (\bibinfo{year}{1997}).

\bibitem[{\citenamefont{Burkard et~al.}(1999)\citenamefont{Burkard, Loss,
  DiVincenzo, and Smolin}}]{burkard:opt_3bit_ecc}
\bibinfo{author}{\bibfnamefont{G.}~\bibnamefont{Burkard}},
  \bibinfo{author}{\bibfnamefont{D.}~\bibnamefont{Loss}},
  \bibinfo{author}{\bibfnamefont{D.~P.} \bibnamefont{DiVincenzo}},
  \bibnamefont{and} \bibinfo{author}{\bibfnamefont{J.~A.}
  \bibnamefont{Smolin}}, \bibinfo{journal}{Phys. Rev. B}
  \textbf{\bibinfo{volume}{60}}, \bibinfo{pages}{11404} (\bibinfo{year}{1999}).

\bibitem[{\citenamefont{Cleve et~al.}(1998)\citenamefont{Cleve, Ekert,
  Macchiavello, and Mosca}}]{cleve:q_alg_revisited}
\bibinfo{author}{\bibfnamefont{R.}~\bibnamefont{Cleve}},
  \bibinfo{author}{\bibfnamefont{A.}~\bibnamefont{Ekert}},
  \bibinfo{author}{\bibfnamefont{C.}~\bibnamefont{Macchiavello}},
  \bibnamefont{and} \bibinfo{author}{\bibfnamefont{M.}~\bibnamefont{Mosca}},
  \bibinfo{journal}{Proc.\ R.\ Soc.\ London A} \textbf{\bibinfo{volume}{454}},
  \bibinfo{pages}{339} (\bibinfo{year}{1998}).

\bibitem[{\citenamefont{Draper}(2000)}]{draper:qc_adder}
\bibinfo{author}{\bibfnamefont{T.~G.} \bibnamefont{Draper}}
  (\bibinfo{year}{2000}), \eprint{quant-ph/0008033}.

\bibitem[{\citenamefont{Beauregard}(2002)}]{beauregard:shor_2n3}
\bibinfo{author}{\bibfnamefont{S.}~\bibnamefont{Beauregard}}
  (\bibinfo{year}{2002}), \eprint{quant-ph/0205095}.

\end{thebibliography}

\end{document}